\documentclass[prl,aps,twocolumn,nofootinbib,preprintnumbers,superscriptaddress]{revtex4-2}

\usepackage{amsmath}
\usepackage{color}
\usepackage{graphicx}
\usepackage{hyperref}
\hypersetup{colorlinks, citecolor=blue, linkcolor=black, urlcolor=blue}
\usepackage[caption=false]{subfig}

\usepackage[american]{babel}
\usepackage{microtype}

\definecolor{darkblue}{rgb}{0.0,0.0,0.6}

\newcommand{\rhoDM}{\rho_{_{\text{DM}}}}
\newcommand{\mDM}{M_{\rm DM}}
\newcommand{\rDM}{R_{\rm DM}}
\newcommand{\Sun}{\odot}

\usepackage{geometry}
\usepackage{fullpage}
\begin{document}

\title{Stellar Shocks From Dark Matter Asteroid Impacts}

\author{Anirban~Das}
\email{anirband@slac.stanford.edu}
\affiliation{SLAC National Accelerator Laboratory, 2575 Sand Hill Road, Menlo Park, CA 94025, USA}
\author{Sebastian~A.~R.~Ellis}
\email{sarellis@protonmail.com}
\affiliation{Institut de Physique Th\'eorique, Universit\'e Paris Saclay, CEA, F-91191 Gif-sur-Yvette, France}
\affiliation{SLAC National Accelerator Laboratory, 2575 Sand Hill Road, Menlo Park, CA 94025, USA}
\author{Philip~C.~Schuster}
\email{schuster@slac.stanford.edu}
\affiliation{SLAC National Accelerator Laboratory, 2575 Sand Hill Road, Menlo Park, CA 94025, USA}
\author{Kevin Zhou}
\email{knzhou@stanford.edu}
\affiliation{SLAC National Accelerator Laboratory, 2575 Sand Hill Road, Menlo Park, CA 94025, USA}

\preprint{SLAC-PUB-17604}
\begin{abstract}
Macroscopic dark matter is almost unconstrained over a wide ``asteroid-like'' mass range, where it could scatter on baryonic matter with geometric cross section. We show that when such an object travels through a star, it produces shock waves which reach the stellar surface, leading to a distinctive transient optical, UV and X-ray emission. This signature can be searched for on a variety of stellar types and locations. In a dense globular cluster, such events occur far more often than flare backgrounds, and an existing UV telescope could probe orders of magnitude in dark matter mass in one week of dedicated observation. 
\end{abstract}

\maketitle

Astronomical and cosmological observations have provided all evidence for dark matter (DM) thus far. Stars, substellar objects, and stellar remnants therefore comprise natural venues for probing the nature of DM. The capture of particle DM is well-studied, and can result in signatures ranging from heating~\cite{bertone2008compact,kouvaris2008wimp,Baryakhtar:2017dbj,PhysRevLett.126.161101} and modifications of stellar structure~\cite{iocco2012main,casanellas2013first}, to outright destruction by the formation of a black hole~\cite{goldman1989weakly,Acevedo:2020gro,Dasgupta:2020mqg}. DM candidates light enough to be produced thermally in stars, such as the axion, can also be constrained by stellar cooling rates~\cite{raffelt1996stars}. 

However, DM could also be in the form of objects of macroscopic mass and size, a possibility which is consistent with all cosmological constraints~\cite{jacobs2015macro,burdin2015non,Bai:2020jfm}. While macroscopic DM arises in many theoretical scenarios, it is difficult to detect terrestrially primarily because such objects are rare, given the low local DM density. As $\mDM$ increases, experimental searches require either large detection volumes or long integration times. For example, for $\mDM \lesssim 10^5 \, \mathrm{kg}$, limits on macroscopic DM passing near the Earth can be set with tabletop experiments, calorimeters, and gravitational wave detectors~\cite{Grabowska:2018lnd,budker2020axion}, or searches for fast-moving meteors~\cite{sidhu2019macro,sidhu2019macroscopic,piotrowski2020limits} and seismic waves~\cite{herrin2006seismic,cyncynates2017reconsidering}. However, for $\mDM \gtrsim 10^{-20} M_\Sun$, corresponding to a heavy asteroid, macroscopic DM would not have collided with Earth since the advent of human civilization, and setting constraints requires speculative appeals to geologic history~\cite{abbas1998volcanogenic,rafelski2013compact2}. Unambiguously probing the mass range  $10^{-20} M_\Sun \lesssim \mDM \lesssim 10^{-11} M_\Sun$ of ``dark asteroids'' will therefore require looking to the stars. 

In this Letter, we point out that because dark asteroids move supersonically in stars, dissipation through any non-gravitational interaction will generate shock waves. This allows the dissipated energy to quickly propagate to the stellar surface, where it is released in the form of a transient, thermal ultraviolet (UV) emission. Crucially, such events are correlated with the local DM density, but uncorrelated with the underlying activity of the star. Next-generation survey telescopes would detect such events without requiring a dedicated search, while existing telescopes could find them by monitoring regions of high DM density. This would constitute a DM direct detection experiment on astronomical scales, with the stars as the detector volume.

A detailed overview of models that produce dark asteroids is beyond the scope of this work, but the reader can keep several scenarios in mind. Self-interactions in the dark sector allow models as simple as asymmetric DM~\cite{Zurek:2013wia} to build up composite objects of high multiplicity in the early universe~\cite{Hardy:2014mqa,Gresham:2017cvl}, and support compact structures~\cite{Kouvaris:2015rea,Gresham:2018anj,Chang:2018bgx}. Additionally introducing a lighter, oppositely-charged particle allows dark atoms to form, providing an alternative cooling mechanism which can generate large DM structures~\cite{Buckley:2017ttd}, while charging the DM under a non-Abelian gauge group naturally allows dark nucleosynthesis~\cite{Krnjaic:2014xza,Detmold:2014qqa}. An even richer dark sector, which could result from mirroring part or all of the Standard Model, allows the formation of mirror stars~\cite{foot2004mirror,berezhiani2006evolutionary,DAmico:2017lqj,curtin2020signatures}. Phase transitions in the dark sector can also produce large dark objects, for both bosonic and fermionic DM~\cite{witten1984cosmic,Frieman:1988ut,bai2019dark,gross2021dark}, with the density determined by the temperature of the phase transition. 

For concreteness, we will introduce our signature by assuming that all DM is in the form of spherical dark asteroids with the same mass $\mDM$ and radius $\rDM$. We further assume that they scatter baryons elastically with geometric cross section $\sigma = \pi \rDM^2$, enter the star head-on, and do not disintegrate while passing through the star. In the final section, we discuss how these properties can arise and how the signature changes when they are relaxed. 


\textit{Stellar collisions.}---We compute stellar profiles with MESA~\cite{Paxton2011,Paxton2013,Paxton2015,Paxton2018,Paxton2019}, assuming solar metallicity and the settings recommended by MIST~\cite{choi2016mesa}, and match them at the photosphere to atmospheric profiles computed with PHOENIX~\cite{husser2013new}. When a dark asteroid enters a star of mass $M_\star$ and radius $R_\star$, it will be traveling at roughly the escape velocity $v_{\mathrm{esc}} = \sqrt{2 G M_\star / R_\star}$, and is therefore hypersonic, with Mach number $\mathrm{Ma} \sim 100$. It is accelerated inward by gravity, and dissipates energy due to a drag force $\rho \sigma v^2 c_d/2$, where $c_d \simeq 1$ for a supersonic sphere~\cite{bailey1972sphere}. For most of the parameters we consider, the dark asteroid remains hypersonic until it either dissipates most of its energy to drag, or reaches the hot stellar core. 

Describing the resulting production and propagation of shock waves is a complex hydrodynamic problem. However, it can be decomposed into simpler problems each solvable by controlled approximations, as shown in more detail in the Supplemental Material. First, because the dark asteroid is hypersonic, $\mathrm{Ma} \gg 1$, its passage can be treated as an instantaneous deposition of energy $F_{\mathrm{dr}}$ per unit length, which creates a cylindrical blast wave. Numeric blast wave solutions are known, and are used to model meteors traversing the Earth's atmosphere~\cite{revelle1976meteor,silber2014optical}. The shock wave becomes weak after it travels a characteristic radial distance $R_0 = \sqrt{2 F_{\mathrm{dr}} / p} \sim \mathrm{Ma} \, \rDM$, and asymptotically approaches an N-wave profile, a weak shock solution characterized by a pressure discontinuity $\Delta p$ and length $L$. Following Ref.~\cite{revelle1976meteor}, we match a blast wave onto an N-wave profile at distance $10 R_0$, where the shock strength is $\Delta p / p = 0.06$, the length is $L = 2.8 R_0$, and roughly half of the original energy remains in the shock wave.

To treat the propagation to the stellar surface, we use standard results from weak shock theory~\cite{whitham1956propagation,pierce2019acoustics}. In particular, the propagation of a weak shock wave through a slowly varying medium can be described by geometric acoustics. Because the speed of sound decreases with distance from the center of the star, the ray paths refract radially outward. We propagate each piece of the shock front along such a ray. For an acoustic wave, if the wavefront area evolves as $A(s)$ along a ray, then the pressure amplitude varies as $\Delta p \propto \sqrt{\rho c_s / A(s)}$, while the period $L / c_s$ remains constant. The discontinuities of an N-wave cause additional dissipation: when the shock wave travels a length $L$, there is a fractional increase in $L$, and a fractional decrease in shock strength and total energy, of order $\Delta p / p$.


Finally, as each piece of the shock front approaches the stellar surface, the decreasing density and pressure cause a rapid increase in the shock strength. For the DM masses and radii of interest here, the shock becomes strong, $\Delta p / p \gtrsim 1$, below the photosphere, at optical depths up to $\sim 10^2$. Analytic solutions exist to describe the arrival of a strong shock wave at the edge of a star~\cite{sakurai1960problem}. To roughly approximate these results, we assume that once the shock wave becomes strong, its remaining energy heats the stellar material above it to a uniform temperature $T_f$, which sets the typical frequency band of emission. This is reasonable because convection in the shocked region near the stellar surface will effectively smooth out temperature gradients. The timescale for energy release is then dictated by the rate of blackbody radiation from the surface, and is typically on the order of hundreds of seconds.

This treatment is compatible with previous work on shock waves in stars. In massive stars just prior to core collapse, convection can excite acoustic waves~\cite{quataert2012wave} which then steepen into weak shocks, which dissipate in the same way as they travel outward~\cite{Ro:2016jyo,fuller2018pre}. Refs.~\cite{Matzner_2021,linial2021partial} considered the strengthening of a shock wave near the surface of a star; consistent with this work, we find that our shocks are insufficiently energetic to eject mass, as they emerge with a typical speed $\sqrt{k_B T_f / m_p} \ll v_{\mathrm{esc}}$. 

\begin{figure}[t]
\includegraphics[width=\columnwidth]{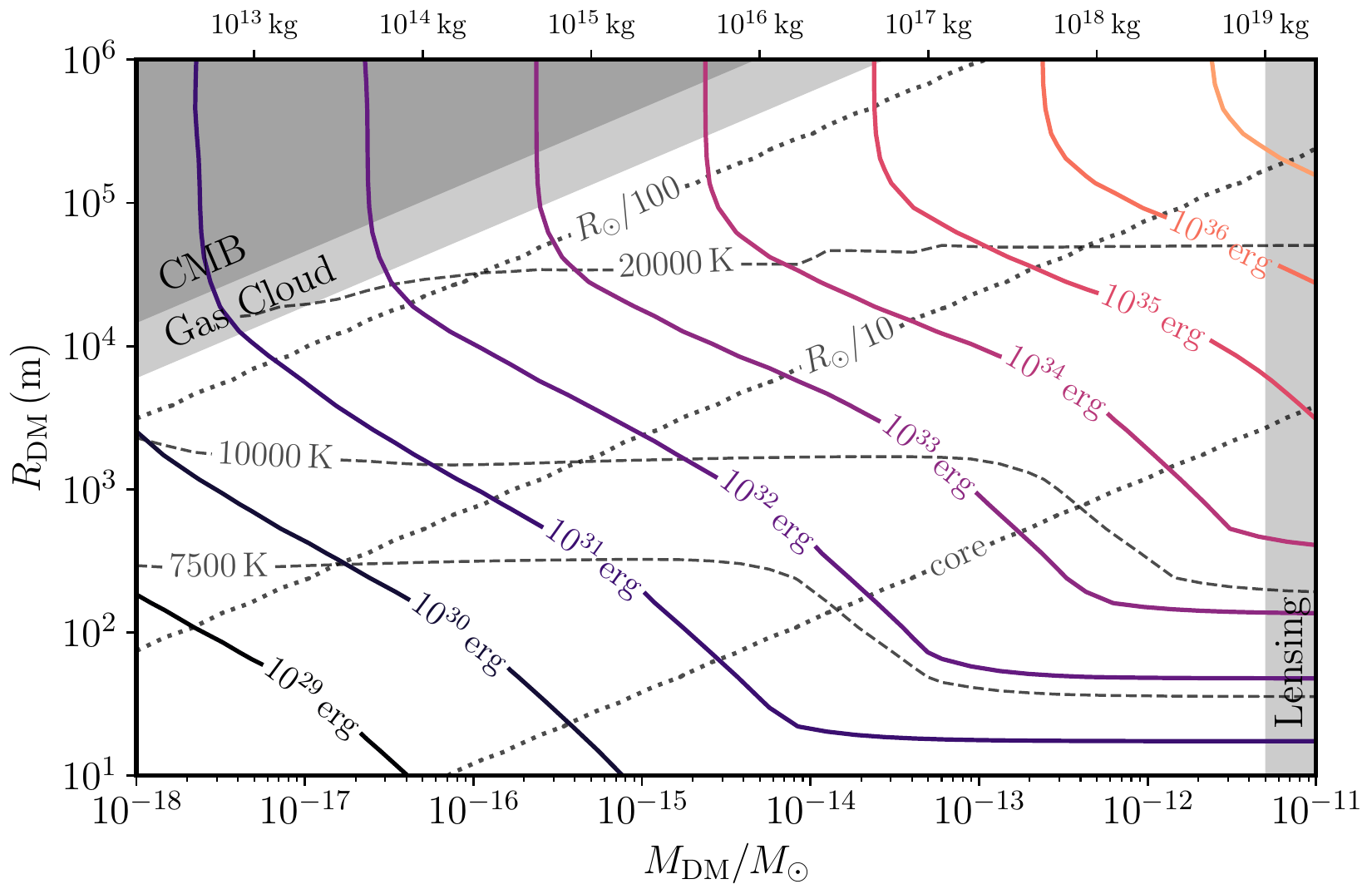}
\caption{Contours of energy release (solid), characteristic temperature (dashed), and penetration depth (dotted) for a dark asteroid impact on a Sun-like star. We show bounds from the CMB limit on DM-baryon scattering~\cite{dvorkin2014constraining}, heating of cold gas clouds~\cite{Bhoonah:2020dzs} (though see also Ref.~\cite{Wadekar:2019xnf}), and microlensing~\cite{niikura2019microlensing} (though see also Refs.~\cite{montero2019revisiting,smyth2020updated,Croon:2020ouk}). We do not show constraints from femtolensing of gamma-ray bursts~\cite{barnacka2012new}, which are weakened by finite source size effects~\cite{katz2018femtolensing}.}
\label{fig:energy}
\end{figure}

In Fig.~\ref{fig:energy}, we show the total shock energy released from the surface of a Sun-like star, and the typical final temperature $T_f$. The qualitative features of this plot can be readily understood. For higher $\rDM$, the dark asteroid stops near the stellar surface, and a small portion of the surface is heated to a high temperature. As $\rDM$ decreases, the shock waves are primarily produced deeper in the star, with a shorter wavelength. This increases the dissipation they experience as they propagate out to the surface, decreasing the energy released. At the smallest radii, drag is insufficient to prevent the dark asteroid from passing through the entire star, so that only part of its energy is deposited, leading to a rapid fall-off in signal energy.

Since a strong shock has $\Delta T / T \sim 1$, the temperature $T_f$ roughly tracks the local temperature at the depth where the weak shock becomes strong again; as a result, it is relatively insensitive to $\mDM$ and $\rDM$, and typically peaks in the far UV. At lower densities, $T_f$ rapidly rises because the dark asteroid stops so close to the surface that the shock never becomes weak. At the very lowest densities shown, the dark asteroid stops above the photosphere. In this extreme case, the emission spectrum is not necessarily thermal, and depends on the detailed physics of the resulting plasma. We do not study this regime because it is in tension with cosmological constraints, but we expect photons to be released at up to X-ray energies, $\epsilon \sim m_p v_{\mathrm{esc}}^2 \sim 10^4 \, \mathrm{eV}$.

Similar results apply to other star types, and are shown in the Supplemental Material. The main difference is that for equal $\mDM$ and $\rDM$, the signal energy is higher for more compact objects, such as red and brown dwarfs, because their density profiles rise more steeply with depth, causing the dark asteroid's energy to be deposited closer to the surface. Conversely, the signal energy is significantly lower for giant stars, because of their extended envelopes.

\begin{figure}[t]
\includegraphics[width=\columnwidth]{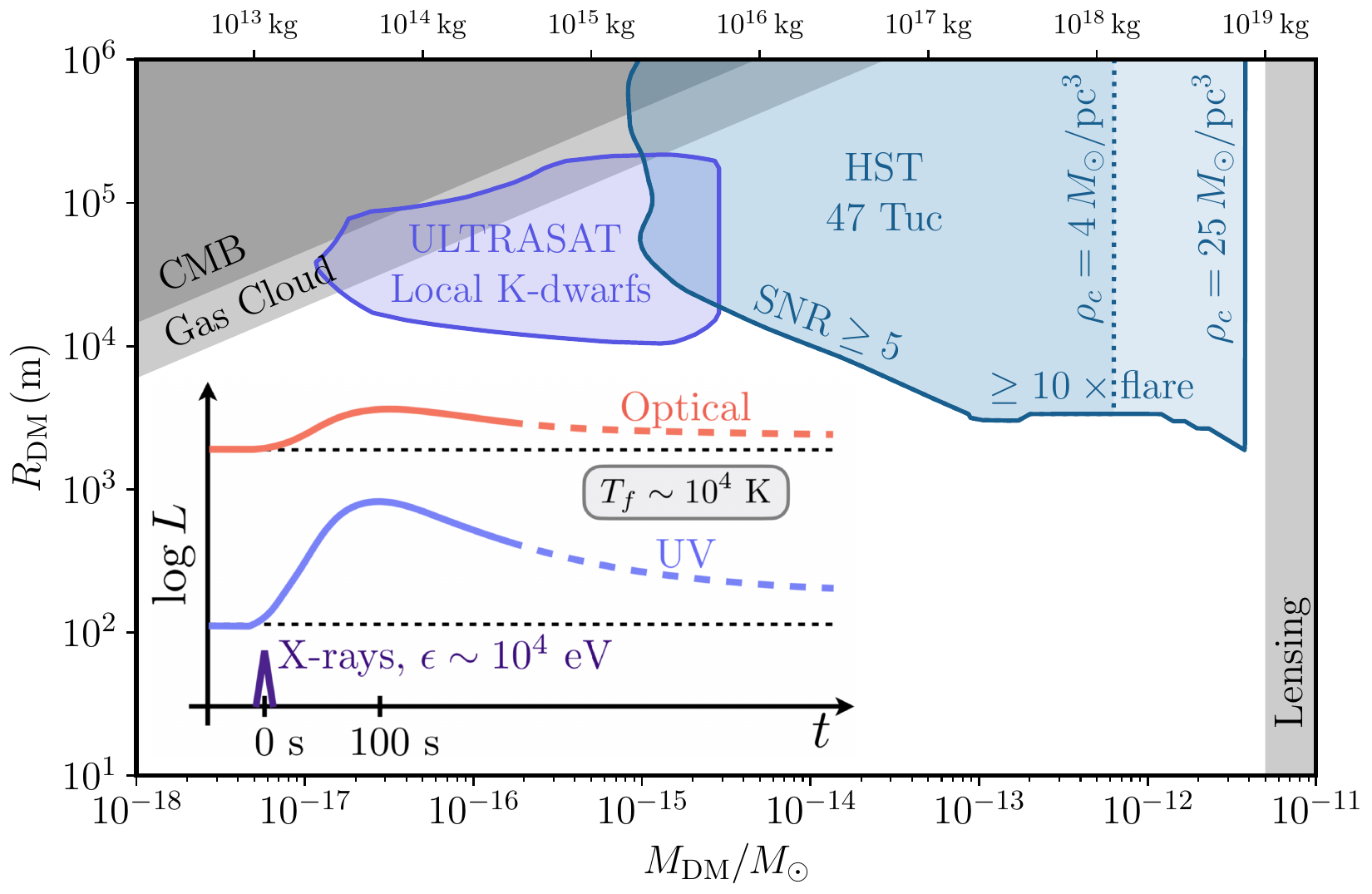}
\caption{Contour plot showing observability. We shade regions where impacts on K dwarfs within $1 \, \mathrm{kpc}$ would be seen by ULTRASAT, and impacts on Sun-like stars in 47 Tuc would be seen by HST, at least once per year and week of observation on average, respectively. In both cases, we demand $\text{SNR} \geq 5$. For 47 Tuc, we show two possible values of the core DM density, as discussed in the main text, and require the rate of dark asteroid impacts to exceed superflares of similar energy by at least an order of magnitude. A schematic light curve for three frequency bands is shown in the inset.}
\label{fig:rate}
\end{figure}

\textit{Observational prospects.}---Dark asteroids are expected to produce rare transients on all types of stars, with a frequency dependent on the stellar and local DM parameters. For a star moving with a DM halo, averaging over a Maxwellian velocity distribution for the DM yields a collision rate~\cite{gould1987resonant}
\begin{equation}
\Gamma = \sqrt{\frac{8}{3\pi}} \frac{\rhoDM v_d}{\mDM} \, \pi R_{\star}^2  \left( 1 + \frac{3 v_{\mathrm{esc}}^2}{2 v_d^2} \right)
\label{eq:rateEq}
\end{equation}
where $v_d$ is the velocity dispersion. The final term accounts for the focusing effect of gravitational attraction. In all cases we will consider, $v_{\text{esc}} \gg v_d$, giving
\begin{multline}
\Gamma \simeq (4 \times 10^{-5} \, \mathrm{yr}^{-1}) \, \frac{M_\star}{M_\Sun} \frac{R_\star}{R_\Sun} \\ \times \frac{10^{-15} M_\Sun}{\mDM} \frac{\rhoDM}{0.4 \, \mathrm{GeV}/\mathrm{cm}^3} \frac{270 \, \mathrm{km/s}}{v_d}.
\label{eq:rateSol}
\end{multline}
As shown in the inset of figure~\ref{fig:rate}, we expect a brief X-ray emission as the dark asteroid passes through the stellar atmosphere, followed by a gradual optical and UV emission as the shock wave produced inside reaches the surface of the star. Since most of the energy emerges in the UV, and cooler stars emit relatively little in this band, it is easiest to search for these events as UV transients. 

The light curve would also have a long tail as the violently heated patch of the stellar surface gradually cools, which could be targeted for follow-up optical observation. Note that we have treated all collisions as head-on, though the high degree of gravitational focusing implies that most collisions are glancing. Our calculation is thus maximally conservative, because it gives the shock waves the longest possible path to the surface. Accounting for the impact parameters would increase the signal strength and temperature, and could also increase visibility to X-ray telescopes.

Upcoming transient surveys could detect dark asteroid collisions on nearby stars without requiring a dedicated search. Among star types, K dwarfs are promising targets, as they have significantly larger masses and radii than M dwarfs, but also have a higher number density and negligible UV emission compared to hotter stars. As a concrete example, we consider ULTRASAT~\cite{sagiv2014science}, a proposed wide-field UV transient explorer designed to detect distant supernova shock breakouts, which will also monitor many nearby stars. We compute the maximum distance from which ULTRASAT could observe dark asteroid collisions at $\text{SNR} \geq 5$, conservatively counting only impacts on K dwarfs, and approximate the star density as uniform out to $1 \, \mathrm{kpc}$ from the Earth. The observable region of parameter space is cut off at high $\rDM$ because the signal temperature becomes too high, at low $\rDM$ and $\mDM$ because the signal energy becomes too low, and at high $\mDM$ because the events become too rare. 

A similar region could be probed by the upcoming LSST survey~\cite{ivezic2019lsst}, but estimating the event rate is more difficult because of LSST's complex observing strategy and multiple filters. In addition, since LSST would be able to see events at a significantly larger distance $d \gtrsim \text{kpc}$, a more detailed model of the galactic stellar and DM densities would be required, along with estimates of UV extinction. Exoplanet searches such as TESS~\cite{ricker2014transiting} and the planned PLATO mission~\cite{rauer2013plato} have exceptionally large fields of view, but observe in the red, which reduces the sensitivity because of stellar variability and shot noise. However, these instruments could effectively detect transients on cool red dwarfs or brown dwarfs, which thereby probes lower $\rDM$, as shown in the Supplemental Material. 

Because the local DM density is low, a potential obstacle for any local search is the background from stellar superflares, which occupy a similar temperature and energy range. Observations from Kepler~\cite{Davenport_2016,okamoto2021statistical} and TESS~\cite{2020AJ....159...60G} find no superflares on the vast majority of FGK dwarfs, and almost none on those that are not rotating rapidly, which allows highly active stars to be excluded from observation. However, for any individual event, it would be difficult to rule out the possibility of a superflare without further information. For example, follow-up observations could determine the detailed light curve, which could fall off more slowly for dark asteroid impacts because the energy emerges from within the star rather than from its atmosphere. Simultaneous observation with other instruments could rule out flares using spectral information, as they are expected to have a significant radio and X-ray component.

An alternative strategy is to perform a focused search in a region where the impact rate per star is significantly higher. As a concrete example, we consider 47 Tuc (NGC 104), a well-studied nearby ($d \sim 4 \, \mathrm{kpc}$) globular cluster which has a dense core and negligible UV dust extinction~\cite{schlafly2011measuring}. 

While the DM content of globular clusters today is not known~\cite{moore1996constraints,Ibata:2012eq}, they are thought to have formed in large DM subhalos~\cite{Peebles:1984zz,Creasey:2018bgv}, with computational studies suggesting an initial DM mass of about $260$ times the stellar mass~\cite{Griffen_2010}. Tidal stripping and DM thermalization are expected to have reduced the DM content of the globular cluster since formation~\cite{Saitoh:2005tt}, with $\sim\!1\%$ remaining today~\cite{Gao_2004}. We assume this formation history holds for 47 Tuc, and model the DM distribution with an NFW profile~\cite{Navarro_1996}. Gravitational interactions transfer kinetic energy to the dark asteroids and lighter stars, which we account for by coring the DM profile~\cite{Merritt_2004} and taking a relatively heavy benchmark star of solar mass.

From the above procedure, detailed in the Supplemental Material, we infer a core DM density $\rhoDM \simeq 4 \, M_\Sun / \mathrm{pc}^3$. Since the velocity dispersion is $v_d \simeq 12 \, \mathrm{km}/\mathrm{s}$, the collision rate per star is almost $4$ orders of magnitude higher than in the local region, even though DM is still a vastly subdominant component of the core. For most of the parameters we consider, the event rate exceeds the rate of superflares of comparable energy on Sun-like stars~\cite{okamoto2021statistical} by orders of magnitude. Yet our estimate is conservative, as a recent analysis with similar assumptions~\cite{bertone2008compact} found a core DM density $\rhoDM \simeq 25 \, M_\Sun / \mathrm{pc}^3$~\cite{Leane:2021ihh}. Furthermore, we neglect adiabatic contraction of the DM halo~\cite{Blumenthal:1985qy,Gnedin:2003rj}, which would significantly increase the core density, and we do not consider the possibility of a DM cusp due to an intermediate-mass black hole~\cite{brown2018understanding,amaro2016probing}, within which the DM density would be enhanced by orders of magnitude. 

To monitor 47 Tuc, we consider the Wide Field Camera 3 instrument on the Hubble Space Telescope (HST), using the F225W filter. This instrument's field of view is sufficient to capture most of the DM core, and the UV filter alleviates stellar crowding~\cite{knigge2002far}. In Fig.~\ref{fig:rate}, we show the region where at least one event with $\text{SNR} \geq 5$ is expected in one week of continuous observation. HST has in fact already monitored 47 Tuc for over a week to search for exoplanets~\cite{gilliland2000lack}, though these optical and infrared observations are less useful for our purposes due to stellar backgrounds. Since the event rate scales as $1/\mDM$, new parameter space could be probed with as little as one hour of dedicated UV observation.


\textit{Discussion.}---For concreteness, we have focused on specific assumptions and experimental searches, but our results also apply more generally. For instance, we have taken elastic scattering as a generic benchmark, but specific models can give rise to nonelastic interactions, such as catalyzing proton decay, annihilating with ordinary matter, or absorbing part of the dissipated energy. We have also assumed a geometric cross section for baryon scattering because it is the result of any sufficiently strong interaction that is not long-ranged, but the dark asteroid can be partly transparent to baryons, or interact by a long-range force, yielding a smaller or larger cross section respectively. These effects can be accounted for by simply scaling the energy deposited per length, $F_{\mathrm{dr}}$, as long as $R_0 \gtrsim \rDM$. 

The assumption of geometric cross section implies relatively strong DM interactions with the SM, and it is interesting to see how this can be compatible with existing constraints. As shown in Fig.~\ref{fig:energy}, cosmological constraints are relatively weak, essentially because dark asteroids would be extremely rare, and the constraint from DM self-interaction in the Bullet cluster is orders of magnitude weaker. Strong DM constituent interactions with the SM could be accommodated if, for example, the dark asteroid was composed of ``dark atoms" comprised of oppositely charged particles bound by a dark $U(1)$. If the constituent has $m_{\rm DM} \lesssim 100~\text{MeV}$, and less than $1\%$ is unbound, terrestrial constraints require that the DM-nucleon cross section not exceed $\sigma_{\rm SI} \lesssim 10^{-29}~\text{cm}^2$~\cite{Bringmann_2019,Alkhatib:2020slm}. A dark asteroid of density $\text{g/cm}^3$ containing $100~\text{MeV}$ constituents with this SM interaction strength would have a mean free path of $\sim\! 100~\text{m}$, so that the assumption of opacity holds for the regions of interest of Fig.~\ref{fig:rate}.

Our rough estimates of the collision rate and signal energy could be refined in many ways. We have taken all dark asteroids to have the same mass $\mDM$, though a realistic production mechanism would lead to a range of masses. This would not necessarily harm prospects for a local search, as impacts of heavier dark asteroids could be seen from further away. Dark asteroids could pass through a star but lose sufficient energy to be captured, ensuring subsequent collisions. This enhancement of the event rate occurs for a wide range of $\mDM$ because most collisions are glancing, and leads to the intriguing possibility of followup detection. Finally, our treatment of the shock propagation and convective energy transport near the surface could be improved with detailed analytic arguments or numerics, which would also yield the detailed shape of the light curve.

Collision events have been investigated in related contexts, though the results are qualitatively different. Primordial black holes passing through stars deposit a small amount of energy through dynamical friction, but the result is too weak to observe~\cite{abramowicz2009no,kesden2011transient}. Dark asteroids could also trigger supernova in white dwarfs by depositing energy in their interiors~\cite{graham2018white,sidhu2020reconsidering}. The survival of white dwarfs therefore implies a strong constraint on macroscopic DM due to the long effective integration time, but it only applies to dark asteroids of roughly nuclear density, which can penetrate the white dwarf's crust. 

Within the Standard Model, the closest analogue to a dark asteroid impact would be a comet impact~\cite{jones2018science}. However, comets are expected to be rare outside of planetary systems, with the interstellar comet density bounded orders of magnitude below the DM density~\cite{francis2005demographics}. Comets are also ``rubble piles'' which fall apart before even reaching the stellar surface, leading to a qualitatively different signature. By contrast, in simple dark sector models the binding energy of a dark asteroid may easily exceed its kinetic energy, which is only about $(v_d/c)^2 \sim 10^{-6}$ of its total mass energy, implying that ablation is a small effect.

Many additional directions could be explored in future work. For instance, the high DM density at the galactic center would make it ideal for a focused search, though one would have to model its distinct stellar populations and use a sightline with low extinction. Globular clusters besides 47 Tuc could be promising, especially if new nearby clusters are found, or confirmed to contain an intermediate-mass black hole. Milky Way satellite galaxies are more distant but are known to host a high DM density, and could likely be used to probe higher $\mDM$. At the opposite end of the mass range, impacts on the Sun are expected to occur annually for $\mDM \lesssim 10^{-19} M_\Sun$, and would be energetic enough to be easily detected by solar observatories. It would be interesting to see if the resolution of these instruments permits such impacts to be distinguished from solar flares. In many of these cases, it may be possible to find impact events in a reanalysis of archival data.

The possibility of detecting dark asteroid impacts in nearby stars provides an interesting target for UV transient searches with small satellites~\cite{brosch2014small,mathew2018wide,serjeant2020future}, while more powerful instruments would be well-suited for focused searches. These observations are enabled by the rapid advance of time-domain astronomy, which we have shown provides an unusual route to discovering the nature of dark matter. 

\textit{Acknowledgments.}---We thank Patrick Eggenberger, Rebecca Leane, Bruce Macintosh, Eric Mamajek, Georges Meynet, Payel Mukhopadhyay, Stephen Ro, Ningqiang Song, and Natalia Toro for helpful discussions. AD, SARE, PS, and KZ were supported by the U.S. Department of Energy under contract number DE-AC02-76SF00515 while at SLAC. SARE was also partially supported by SNF grant P400P2\_186678. KZ is supported by the NSF GRFP under grant DGE-1656518.

\bibliographystyle{utphys}
\bibliography{DMStar}


\clearpage
\appendix
\onecolumngrid
\section*{Supplemental Material}

\subsection*{Details of shock calculation}

\begin{figure}[t]
\centering
\includegraphics[width=0.5\textwidth]{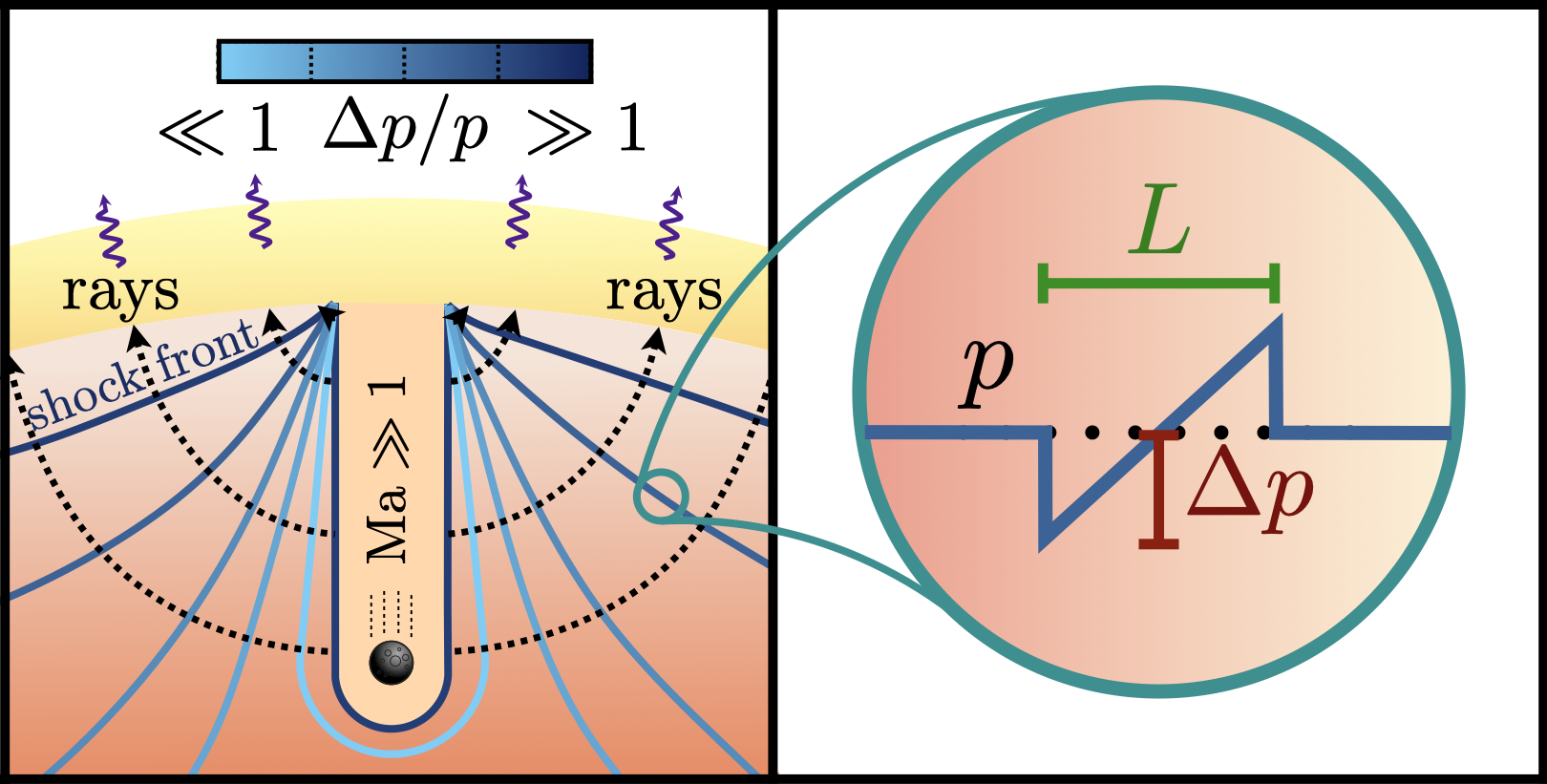}
\caption{Depiction of the phases of shock propagation (left). A cylindrical blast wave solution is matched to an N-wave (right) when the shock become weak. The N-wave is propagated along acoustic rays, and becomes strong and deposits its energy near the surface.}
\label{fig:cartoon}
\end{figure}

In this section, we give further details on the shock wave calculation described in the main body, which is depicted in Fig.~\ref{fig:cartoon}. The initial cylindrical blast wave has been calculated numerically~\cite{plooster1970shock}, and following Ref.~\cite{revelle1976meteor}, its shock strength can be fit as 
\begin{equation}
\frac{\Delta p}{p} \simeq \frac{2 \gamma}{\gamma + 1} \frac{0.4503}{(1 + 4.803 x^3)^{3/8} - 1}
\end{equation}
where $x = r / R_0$, and we set $\gamma = 5/3$. For $x = 10$, the shock wave profile closely approximates an N-wave. The paths of the shock rays for the N-wave are calculated by applying the principle of least time to the sound speed profile $c(r)$ computed by MESA. 

To propagate the N-wave, we use the formulation of Ref.~\cite{pierce2019acoustics}, which states that the pressure profile can be written as 
\begin{equation}
\Delta p(\ell, t) = B(\ell) \, g(\psi)
\end{equation}
where $\ell$ is the arc length along a shock ray. Here, the scaling $B(\ell) \propto \sqrt{\rho c / A(\ell)}$ accounts for energy conservation as the local density, sound speed, and shock front area $A(\ell)$ change, and $B(0) = 1$. The pressure profile is a function of 
\begin{equation}
\psi = t - \tau(\ell) + g(\psi) \mathcal{A}(\ell)
\end{equation}
where $\tau$ is the travel time, $d\tau = d \ell / c$. In the absence of nonlinear effects, $\psi = t - \tau$, indicating that the wave maintains its temporal profile. The shock age $\mathcal{A}$ accounts for the nonlinearity of the wave, with $d \mathcal{A} = (\beta B / \rho c^3) \, d \ell$ and $\beta = (\gamma + 1)/2$. The pressure profile for an N-wave with initial pressure amplitude $\Delta p_0$ is
\begin{equation}
g(t) = \begin{cases} - \Delta p_0 \, t / T_0 & - T_0 \leq t \leq T_0 \\ 0 & \text{otherwise} \end{cases}
\end{equation}
where $L_0 = 2 c T_0$. Plugging this into the previous results, the pressure amplitude $\Delta p$, length $L$, and total shock wave energy $E$ evolve as $\Delta p = \epsilon B \, \Delta p_0$, $L = L_0 / \epsilon$, and $E = \epsilon E_0$, where 
\begin{equation} \label{eq:epsilon}
\epsilon(\ell) = \frac{1}{\sqrt{1 + (\Delta p_0 / T_0) \mathcal{A}(\ell)}}.
\end{equation}
Dissipation by photon diffusion is also present, but it is subdominant for all of the parameter space probed in Fig.~2 of the main body. 

To calculate the total energy release, we discretize the dark asteroid's trajectory. For each segment, we match to an N-wave, which determines $\Delta p_0$ and $T_0$. This determines the initial N-wave energy for the segment,
\begin{equation}
E_0 \simeq \frac23 \frac{(\Delta p_0)^2}{\rho c^2} (A T_0 c)
\end{equation}
which is approximately half the energy dissipated by the dark asteroid in that segment. We then integrate $\epsilon(\ell)$ along each shock ray until the shock becomes strong again. To calculate the final temperature of each piece of the surface, we take the remaining energy $\epsilon E_0$ of the shock wave and assume it uniformly heats the patch of the surface above it, calculating the heat capacity using the MESA profile. The typical final temperature $T_f$ is defined by an energy-weighted median. 

Two main features of this formalism determine the qualitative behavior of Fig.~1 of the main body. First, the fractional energy dissipation when the shock wave travels a length $L$ is $(\Delta p / p)(\beta / \gamma)$, as can be shown by expanding Eq.~\eqref{eq:epsilon}. (This is equivalent to Eq.~(15) of Ref.~\cite{fuller2018pre} up to a factor of $2$, which arises because Ref.~\cite{fuller2018pre} considers a train of N-waves, while we consider only a single N-wave.) Therefore, there is a significant loss of energy when the shock originates from deep inside the star, cutting off sensitivity to low $\rDM$. Second, the typical initial length $L_0$ is not $\rDM$, as might naively be expected, but the much larger $R_0 \sim \text{Ma} \, \rDM$. This allows a significant fraction of the energy to escape, even though we consider $\rDM$ much smaller than stellar scales. 

For very low $\rDM$, shock dissipation is very effective, so the shock remains weak all the way to the photosphere, and continues traveling outward through the stellar atmosphere. The total energy release is highly suppressed, as indicated at the bottom of Fig.~1 of the main body, but $T_f$ can be much higher due to the high temperatures of the chromosphere and corona. The detection prospects would thus be quite different from the UV signals primarily considered in this work.

For the opposite limit of large $\rDM$ and low dark asteroid densities, the dark asteroid stops very close to the surface, and the stellar profile changes significantly on the scale $R_0$, rendering the cylindrical blast wave solution inapplicable for large $x$. In these cases, we match at $x = 2$ when possible (which we have checked yields almost identical results, in general, to matching at $x = 10$). However, when the shock starts so close to the surface that it never becomes weak, our calculation breaks down entirely, and we simply assume that all the energy escapes, leading to the vertical contours at the top of Fig.~1 of the main body. To very roughly estimate $T_f$ in this case, we assume the volume along the shock rays is heated to a uniform temperature.

\begin{figure}[t]
\centering
\subfloat[][Brown dwarf of mass $0.065 M_\Sun$]{
\includegraphics[width = 0.48\textwidth]{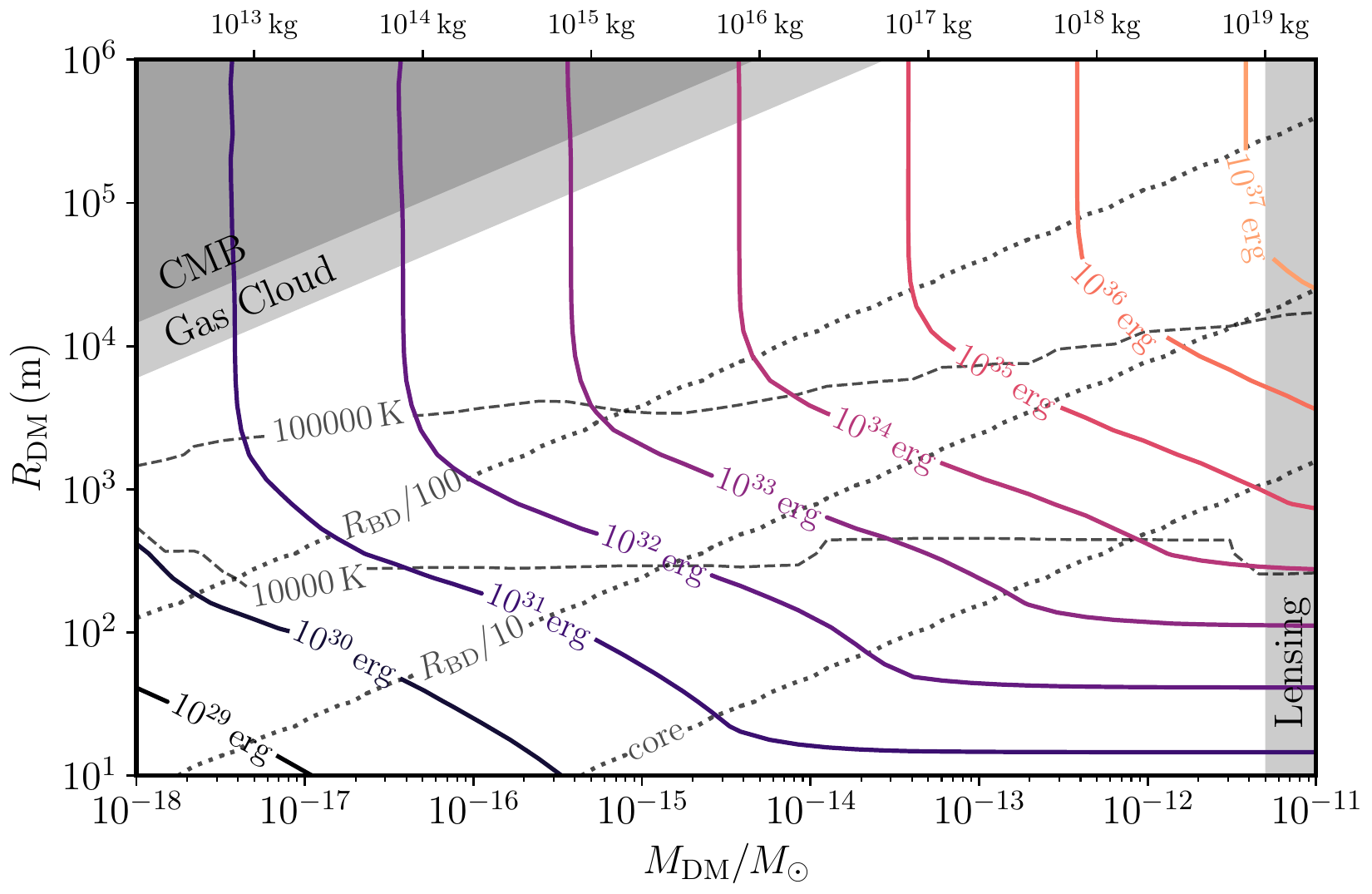}
}
\subfloat[][Red giant of mass $M_\Sun$]{
\includegraphics[width = 0.48\textwidth]{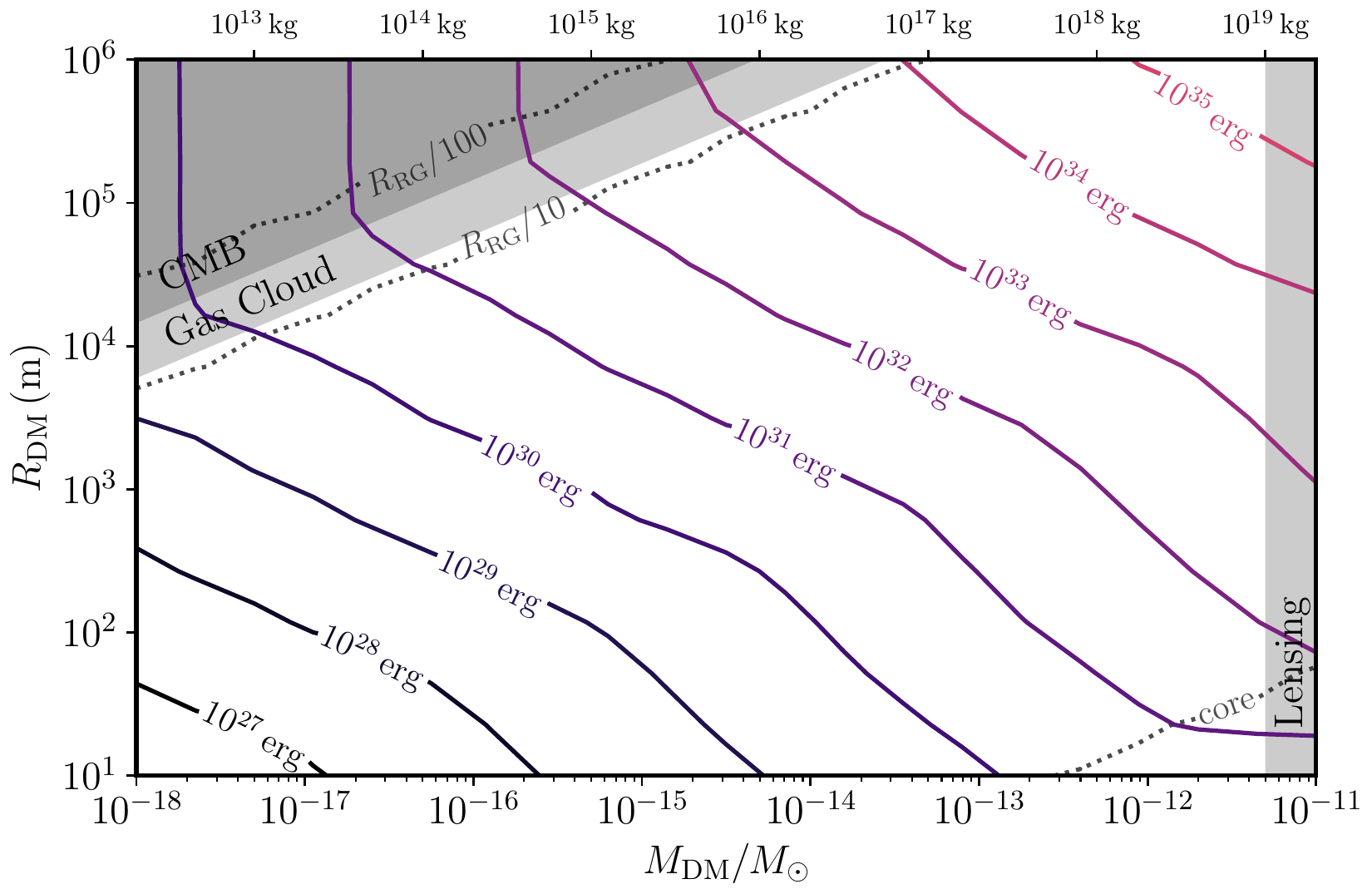}
}
\caption{Analogues of Fig.~1 of the main body for other star types. The differences in signal energy at the lowest dark asteroid density are due to the differences in escape velocity at the surface. Shock dissipation is less severe for brown dwarfs, because dark asteroids are stopped more quickly. It is much more severe for red giants, because even low density dark asteroids can penetrate to a substantial depth. 
}
\label{fig:energy_alt}
\end{figure}

To illustrate these points, we show results for brown dwarfs and red giants in Fig.~\ref{fig:energy_alt}. In red giants, the shock waves lose the vast majority of their energy and remain weak all the way to the surface, where the emission is at temperature $T_f \sim T_\star$. This makes it difficult to observe dark asteroid impacts, even though the rate per star is high. For brown dwarfs, the energy released is high even for very low $\rDM$, but the stellar density profile rises so rapidly that the caveat mentioned in the previous paragraph applies to a wide range $\rDM \gtrsim 1 \, \mathrm{km}$. To get a reliable estimate of $T_f$ in this regime, it may be useful to compare to detailed hydrodynamic studies of comets~\cite{zahnle1994collision} and rocky planets~\cite{anic2007giant} impacting Jupiter. 


\subsection*{Modeling of 47 Tuc}
\label{sec:DensityModels}

\begin{figure}[t]
\includegraphics[width=0.6\textwidth]{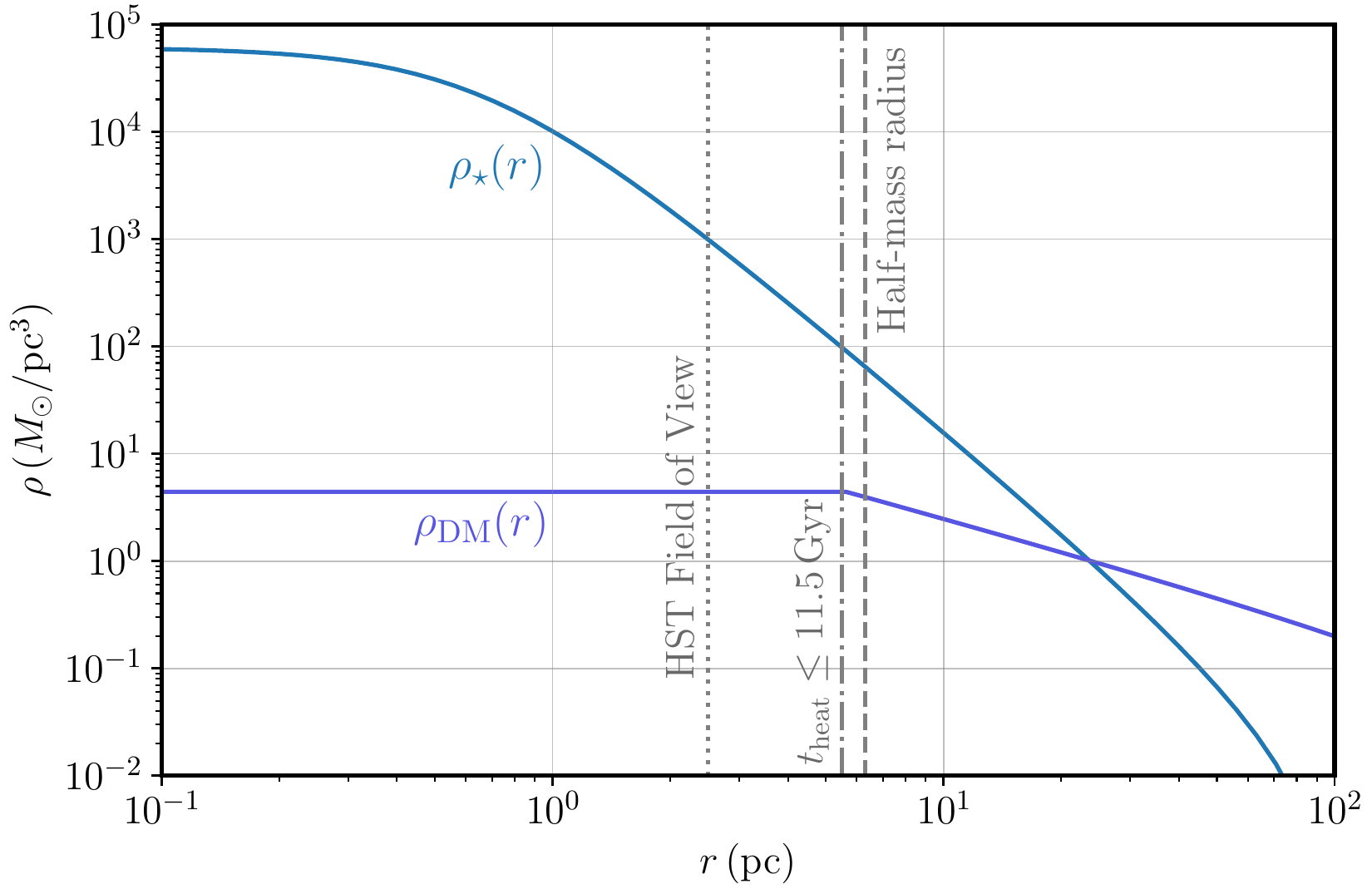}
\caption{Our modeled stellar and DM density profiles for 47 Tuc. We also show the field of view of HST, which encompasses the entire stellar core, but only part of the DM core. 
}
\label{fig:NGC104}
\end{figure}

To find the collision rate between dark asteroids and stars in 47 Tuc, we must estimate the distribution of both species. For the stellar matter, we fit the surface density data reported by Ref.~\cite{Baumgardt_2018} to a 2D King model~\cite{1962AJ.....67..471K},
\begin{equation}
\Sigma(r) = k \left(\frac{1}{\sqrt{1+(r/r_{\rm c})^2}} - \frac{1}{\sqrt{1+(r_{\rm t}/r_{\rm c})^2}} \right)^2,
\end{equation}
where the core radius $r_{\rm c}$ and tidal radius $r_{\rm t}$ are as reported in Ref~\cite{Baumgardt_2018}. Our best fit to the surface density in the GC core yields $k \simeq 6\times 10^4 ~M_\odot/\text{pc}^2$, and the resulting model agrees to within $30\%$ everywhere within $r < 30~\mathrm{pc}$, which is sufficiently accurate for our estimates. Assuming the GC is spherically symmetric, we use the Abel transform,
\begin{equation}
\rho_\star(r) = -\frac{1}{\pi} \int_{r}^{\infty} \frac{d\Sigma(y)}{dy} \frac{dy}{\sqrt{y^2-r^2}} ,
\end{equation}
to obtain the 3D stellar density profile shown in Fig.~\ref{fig:NGC104}.

To model the DM density profile, we assume the globular cluster formed in a DM-rich halo with initial total DM mass $M_{\rm DM, tot} \simeq 260 M_{\star, \rm GC} = 2 \times 10^8 M_\Sun$, as inferred from Ref.~\cite{Griffen_2010}. We model this formation halo with an NFW profile~\cite{Navarro_1996},
\begin{equation}
\rho_{_{\rm NFW}}(r) = \frac{\rho_0}{\left(\frac{r}{a}\right)\left(1 + \frac{r}{a}\right)^{2}} ,
\end{equation}
where $a$ is the scale radius. Here, the reference density is defined in terms of the concentration parameter $c$ as
\begin{equation}
\rho_0 = \frac{M_{\rm DM, tot}}{4\pi a^3} \left(\ln(1+c) -\frac{c}{1+c} \right)^{-1} .
\end{equation}
We infer the values of $a$ and $c$ using the estimated age of 47 Tuc of $11.5 \pm 0.4~\text{Gyr}$~\cite{Campos_2016,Brogaard_2017}, corresponding to a formation redshift $z_f \simeq 3.0 \pm 0.5$ in a $\Lambda$CDM cosmological history. To do this, we set $M_{\rm DM, tot}$ to the halo's virial mass
\begin{equation}
M_{200} = (200 \rho_c(z_f)) \, \frac{4}{3}\pi r_{200}^3,
\end{equation}
defined as the mass corresponding to an average halo density 200 times the critical density of the universe at formation, $\rho_c(z) = 3H(z)^2/8\pi G$. In a $\Lambda$-CDM universe, $H(z) = H_0 \sqrt{\Omega_m(1+z)^3 + \Omega_\Lambda} $, where $H_0 \simeq 67.4~\text{km/s/Mpc}$ today and $\Omega_m = 1-\Omega_\Lambda \simeq 0.315$~\cite{Planck2020}, from which we obtain $r_{200} = 4600 \pm 600~\text{pc}$. The scale radius $a$ is equal to $r_{200} / c$. 

Finally, to compute $c$, we use the semi-analytic results of Ref.~\cite{Ludlow_2016} (which are consistent with the more recent analyses in Refs.~\cite{Diemer_2019,Wang_2020} for our halo mass), which yield $c(z,M_{200}) = 5.8 \pm 0.7$. We therefore choose the parameters $c = 5.8$ and $a = 790 \, \mathrm{pc}$ for the profile shown in Fig.~\ref{fig:NGC104}. Note that the $z_f$ dependence in $c$ and $r_{200}$ roughly cancels, imply little uncertainty in $a$. Therefore, our results are relatively insensitive to uncertainties in the age of 47 Tuc. 

After the globular cluster's formation, the DM is heated through gravitational interactions with stars, which cores the DM profile. Given a typical DM energy $E \sim \mDM v_{\mathrm{rms}}^2$, the rate of change of DM energy due to these interactions is~\cite{Merritt_2004}
\begin{equation}
\frac{dE}{dt} \simeq \frac{8(6\pi)^{1/2}G^2\rho_\star \mDM \ln \Lambda}{\left(v_{\mathrm{rms}}^2 + v_{\star,\mathrm{rms}}^2\right)^{3/2}}(E_\star - E),
\end{equation}
where $\ln \Lambda \sim \ln (0.4 N_\star)$ is the Coulomb logarithm~\cite{1987degc.book.....S} and $N_\star$ is the number of stars in the core of the globular cluster, $v_{\star, \mathrm{rms}}$ is the rms speed of the stars, and $E_*$ is the average stellar kinetic energy. Assuming $v_{\mathrm{rms}} \sim v_{\star, \mathrm{rms}}$ and noting that $E_\star \gg E$, the typical heating timescale is 
\begin{equation}
t_{\rm heat} = \left\vert \frac{1}{E}\frac{dE}{dt}\right\vert^{-1} \sim \frac{0.0814 v_{\rm rms}^3}{G^2 M_\star \rho_\star \ln \Lambda}.
\end{equation}
Setting $t_{\mathrm{heat}}$ to the estimated age of 47 Tuc, $M_\star$ to a solar mass $M_\odot$, and $v_{\mathrm{rms}}$ to the measured velocity dispersion $v_0 = 12.3~\text{km/s}$~\cite{Baumgardt_2018} yields $\rho_\star \simeq 86 \, M_\odot/\mathrm{pc}^3$. We therefore assume the DM profile is cored when the stellar density exceeds this value, which corresponds to coring the NFW profile obtained above at $r_{\mathrm{heat}} \simeq 5.6~\mathrm{pc}$.

As a check that our procedure yields sensible results, integrating the cored NFW profile out to the tidal radius of 47 Tuc gives a total DM mass of $1.9\times 10^6 M_\odot$, which is roughly consistent with the results of the Jeans equation solution models of Ref.~\cite{Ibata:2012eq} for NGC 2419. It should be noted, however, that 47 Tuc, owing to its greater proximity to the galactic center than NGC 2419, might have experienced greater stripping of its DM halo.


\subsection*{Event rate and telescope sensitivity}

We estimate the observable event rate for a search on nearby stars as
\begin{equation}
\Gamma_{\mathrm{tot}} \simeq \frac12 \left( \frac43 \, \pi d_{\mathrm{max}}^3 \right) \frac{\Omega}{4 \pi} \, n_\star \Gamma
\end{equation}
where $d_{\mathrm{max}}$ is the maximum distance at which the events can be observed, $n_\star$ is the stellar density, and $\Omega$ is the angular field of view. The factor of $1/2$ is because impact events can only be seen from one side of the star. We consider K dwarfs with benchmark $M_\star = 0.7 M_\Sun$ and $R_\star = 0.75 R_\Sun$, and following Ref.~\cite{stardensity}, we estimate the local density of K dwarfs to be $n_\star = 0.0135 \, \mathrm{pc}^{-3}$ on the basis of a local census. 

The currently planned field of view of ULTRASAT~\cite{ULTRASATparams} is $\Omega = 200 \, \mathrm{deg}^2$. To find the maximum distance, we assume the signal energy is released isotropically, so that the observed flux at a distance $d$ is 
\begin{equation}
F \simeq \frac{E_{\mathrm{sig}}}{t_{\mathrm{exp}}} \frac{2}{4 \pi d^2}
\end{equation}
where the factor of $2$ is again because events are only visible from the hemisphere of impact, and the timescale is $t_{\mathrm{exp}} = \max(t_{\mathrm{typ}}, 300 \, \mathrm{s})$, to account for ULTRASAT's cadence. We then calculate the AB magnitude of the events in the ULTRASAT band $220$--$280 \, \mathrm{nm}$, assuming a blackbody spectrum of temperature $T_f$. For the events we consider, star noise is subdominant, and the limiting magnitude for detection at $\mathrm{SNR} \geq 5$ given a total exposure time $900 \, \mathrm{s}$ is $22.3$. We scale this sensitivity to a time $t_{\mathrm{exp}}$ and use it to determine $d_{\mathrm{max}}$, which we cap at $1 \, \mathrm{kpc}$.

For a search in 47 Tuc, the event rate is 
\begin{equation}
\Gamma_{\mathrm{tot}} \simeq \int_0^{r_{\mathrm{max}}} dr \, (4 \pi r^2) n_\star \Gamma.
\end{equation}
Because of mass segregation, we take the stars in the core to be Sun-like, $M_* = M_\Sun$ and $R_* = R_\Sun$. Since the distance to 47 Tuc is $d = 4 \, \mathrm{kpc}$, we integrate out to $r_{\mathrm{max}} = 2.5 \, \mathrm{pc}$, accounting for the field of view of the WFC3 instrument of HST. We calculate the AB magnitude of the events in the same way as for ULTRASAT, using the F225W filter, which is most sensitive to wavelengths $210$--$260 \, \mathrm{nm}$. We infer a sensitivity threshold over a timescale $t_{\mathrm{exp}} = \max(t_{\mathrm{typ}}, 300 \, \mathrm{s})$ using the measured sensitivity, at $\text{SNR} = 5$, to sources of magnitude $26.3$ in an exposure time $5400 \, \mathrm{s}$~\cite{windhorst2010hubble}.

Note that because the signal energy varies over orders of magnitude over the parameter space, the region probed is relatively insensitive to order one factors. For instance, taking a higher threshold, such as $\text{SNR} \geq 10$, or removing the factors of $2$ discussed above, would not qualitatively change Fig.~2 of the main body. 

\end{document}